\documentstyle[sprocl,epsfig]{article}

\begin{document}

\title{VECTOR-TO-PSEUDOSCALAR AND MESON-TO-BARYON RATIOS IN HADRONIC Z DECAYS 
AT LEP}

\author{Vladimir UVAROV}

\address{Institute for High Energy Physics, Protvino, Moscow region, Russia\\
E-mail: uvarov@mx.ihep.su}

\maketitle\abstracts{
Mass dependences of the total production rates per hadronic Z decay of all 
light-flavour hadrons measured so far at LEP are extrapolated to the zero mass
limit ($m$=0) using phenomenological laws of hadron production related to the 
spin, isospin, strangeness content and mass of the particles. 
The vector-to-pseudoscalar and meson-to-baryon ratios at $m$=0 are found to be: 
$\rho^+ /\,3\,\pi^+ = 1.2\,\pm\,0.3$ and $\pi^+ /\,{\rm p} = 2.9\,\pm\,0.3$, 
in good agreement with the predictions of quark combinatorics.}

The measurement of the ratios of particle production rates in 
${\rm e}^+{\rm e}^-$
annihilation is fundamental to the understanding of the fragmentation
of quarks and gluons into hadrons. Only phenomenological models, which
need to be tuned to the data, are available to describe this hadronization
process. In particular, the quark combinatorics model~\cite{Anisovich,Bjorken}
predicts the following values for the ratios of the direct production rates of 
vector and pseudoscalar mesons~\cite{Anisovich,Bjorken} and of mesons and 
baryons:\,\cite{Anisovich} 
\begin{equation}
\label{qcmratios}
\rho^+ : \pi^+ = 3 : 1 ~~~~~{\rm and}~~~~~ 
\pi^+ : {\rm p} \,:\, \bar{\rm p} = 3 : 1 : 1.
\end{equation}
These values originate from the usual spin counting factor (2$J$+1) and from 
simple combinatorics of $q\bar{q}$, $qqq$ and $\bar{q}\bar{q}\bar{q}$ 
production. However, the values (\ref{qcmratios}) are not observed 
experimentally in all types of interactions, at least in the central 
kinematical region (see, for example, Refs.\,\cite{ANN,PVratio,HypSpl}). 
The fact that the vector-to-pseudoscalar ratio is suppressed relative to the 
prediction (\ref{qcmratios}) has been known for a long time but is still 
poorly understood. 
The comparative properties of meson and baryon production are not completely 
understood either.

Beyond the cluster fragmentation model~\cite{herwig} and the string 
model,\,\cite{jetset} which employ many parameters to describe hadron 
fragmentation, several attempts were made recently to understand the global
properties of particle production in ${\rm e}^+{\rm e}^-$ annihilation. 
In Ref.\,\cite{StrReg} a similarity in the mass squared dependence of  
meson and baryon production rates was found. It was followed by rather 
successful phenomenological models with very few parameters: 
the ``hadron gas model'',\,\cite{hgm}
the ``string-based model'',\,\cite{sbm}
the ``improved pop-corn model''~\cite{popcorn}
and the ``quark model with constituent quarks''.\,\cite{HypSpl}

The purpose of this analysis, 
presented recently in Refs.\,\cite{MyPLB,Moriond},
is to show that the relations (\ref{qcmratios}) expected from quark 
combinatorics are observed experimentally in hadronic Z decays at LEP 
not for the ratios of the {\it direct} production rates, but for the ratios of 
the {\it ``massless''} particle production rates, i.e. for the ratios obtained 
by extrapolating the empirical mass ($m$) dependences of the {\it total} 
production rates to the zero mass limit $m$=0.

The total\,\footnote{~The quoted rates include decay products from resonances 
and particles with $c\tau < 10$ cm.} production rates per hadronic Z decay
of light-flavour hadrons  used in this analysis were obtained for at least one 
state of a given isomultiplet as a weighted average\,\footnote{~In calculating 
the errors of averages, the standard weighted least-squares procedure 
suggested by the PDG~\cite{PDG} was applied: if the quantity 
$[\chi^2/(N-1)]^{1/2}$ was greater than 1, the error of the average was 
multiplied by this scale factor.} of the measurements of the four LEP 
experiments: ALEPH,\,\cite{Aall} DELPHI,\,\cite{Dall,D9} L3~\cite{Lall} and 
OPAL.\,\cite{Oall} 

\begin{figure}[hbtp]
\centering\mbox{
\epsfig{file=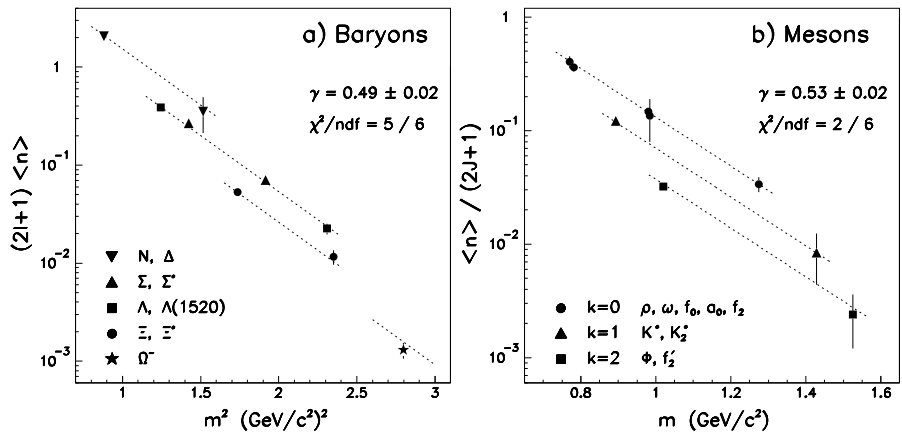,width=\textwidth}}
\caption{a) Sum of the baryon production rates of all states of an 
isomultiplet as a function of the squared baryon mass. b) Meson production 
rate per spin and isospin state as a function of the meson mass.}
\label{fig1}
\end{figure}

It has been shown that the total production rates per hadronic Z decay of 
vector, tensor and scalar mesons~\cite{Panic99} and of baryons~\cite{D9} 
follow phenomenological laws related to the spin ($J$), isospin ($I$), 
strangeness content and mass of the particles. Using the basic idea from 
Refs.\,\cite{Panic99,D9} and adding the latest data measurements, we analyse 
the baryon production rates in a different way from the meson production rates. 
For baryons (Fig.\,\ref{fig1}a), the sum of the production rates of all states 
of an isomultiplet is plotted as a function of $m^2$. In the case where not 
all states of an isomultiplet are measured at LEP, equal production rates for 
the other states is assumed. For mesons (Fig.\,\ref{fig1}b), the production 
rate per spin and isospin state is plotted as a function of $m$. This rate was 
obtained by averaging the rates of particles belonging to the same 
isomultiplet, excluding charge conjugated states and dividing by a spin 
counting factor (2$J$+1).
Therefore, the vertical axes of Fig.\,\ref{fig1}a and Fig.\,\ref{fig1}b
are denoted by $(2I+1)\,{\langle n \rangle}$ and ${\langle n \rangle}/(2J+1)$,
respectively, where $\langle n \rangle$ is the mean production rate
per hadronic Z decay of a given particle.

It has been shown~\cite{D9} that the mass dependence of baryon production 
rates (Fig.\,\ref{fig1}a) is almost identical for the following sets of 
particles: 

\smallskip
\smallskip
1.~~$N$, $\Delta$ with strangeness $S$=0; 

2.~~$\Sigma$, $\Sigma^*$, $\Lambda$ and $\Lambda$(1520) with $\vert S \vert$=1;

3.~~$\Xi$, $\Xi^*$ with $\vert S \vert$=2.

\smallskip
\smallskip
\noindent
Finally the $\Omega^-$ rate is well predicted assuming the same mass dependence
with an additional suppression for the higher strangeness ($\vert S \vert$=3) 
equal to that between the first and second or second and third of the above 
sets.

A similar simple behaviour is seen (Fig.\,\ref{fig1}b) for vector, 
tensor and scalar meson production rates.
The mass dependence of these production rates 
is almost identical for the following sets of particles: 

\smallskip
\smallskip
1.~~$\rho$, $\omega$, f$_0$(980), a$_0$(980), f$_2$(1270); 

2.~~K$^*$, K$^*_2$(1430);

3.~~$\phi$, f$_2^{\,\prime}$(1525).

\smallskip
\smallskip
\noindent
These sets can be considered as sets of mesons with the same number ($k$) of 
$s$ and $\bar{s}$ quarks in the hadron: $k$=0, $k$=1 and $k$=2 
for the 1st, 2nd and 3rd set, respectively. These numbers are close to the well 
known strangeness contents of mesons,\,\cite{PDG} except for the f$_0$(980) 
meson. The interpretation of the latter is one of the most controversial
in meson spectroscopy.\,\cite{AMN} The total production rate of f$_0$(980) 
was measured with an 8\% total error (see compilation in Ref.\,\cite{MyPLB}) 
and in this approach we assume that for the f$_0$(980), $k$=0.

The mass dependences of the total production rates of all light-flavour 
hadrons measured so far at LEP\,1 are well fitted by the following formulae: 
\begin{equation}
\label{blaw}
{\Sigma_i}\,{\Sigma_j}\,{\langle n \rangle}_{ij} \,\equiv\,
(2I+1)\,{\langle n \rangle} \,=\, A\,\gamma^k\,\exp{(-b\,m^2)}
\end{equation}
and
\begin{equation}
\label{mlaw}
{\langle n \rangle}_{ij} \,\equiv\, {\langle n \rangle}\,/\,(2J+1) 
\,=\, A\,\gamma^k\,\exp{(-b\,m)},
\end{equation}
for baryons and mesons, respectively, where ${\langle n \rangle}_{ij}$ is 
the production rate per spin and isospin state. For baryons, 
$k = \vert S \vert$. The values of the fitted parameters $A$, $b$ and $\gamma$ 
are given in Table~\ref{tab2}
\begin{table}[t]
\caption{Values of the fitted parameters $A$, $b$, $\gamma$ and
$\chi^2$ per degree of freedom ($b$ in (GeV/$c^2$)$^{-2}$ for baryons 
and in (GeV/$c^2$)$^{-1}$ for mesons).\label{tab2}}
\vspace{0.2cm}
\begin{center} 
\footnotesize
\begin{tabular}{|ll|c|c|c|c|}
\hline
Particles & & $A$ & $b$ & $\gamma$ & $\chi^2 / ndf$ \\ 
\hline
Baryons & $(B)$  
& 21.4$\pm$1.7 & 2.64$\pm$0.08 & 0.49$\pm$0.02 & 4.6 / 6 \\
Vectors, Tensors, Scalars & $(V,T,S)$ 
& 18.6$\pm$4.3 & 4.95$\pm$0.26 & 0.53$\pm$0.02 & 2.2 / 6 \\
Pseudoscalars & $(P)$ 
& 15.0$\pm$1.2 & 3.85$\pm$0.57 & 0.48$\pm$0.10 & 0.6 / 1 \\
\hline
\end{tabular}
\end{center}
\end{table}
and the fitted curves (\ref{blaw}) and (\ref{mlaw}) are shown in 
Fig.\,\ref{fig1}\,a,b. The result of the fit 
of Eq.~(\ref{mlaw}) to the pseudoscalar meson production rates is also 
given in Table~\ref{tab2} (not shown in Fig.\,\ref{fig1}b). 
The pseudoscalars used are the $\pi$ with $k$=0 and the K, $\eta$, 
$\eta^{\,\prime}$ with $k$=1.
The quark contents ($k$=1) of the $\eta$ and $\eta^{\,\prime}$ mesons originate 
from the singlet-octet mixing angle, $\theta \approx -10^{\circ}$, 
given in Table 13.3 of PDG.\,\cite{PDG}
The strangeness suppression factor $\gamma$ is found to be the same within 
errors for all hadrons (Table~\ref{tab2}), with an average value of 
$\gamma = 0.51 \pm 0.02$, in good agreement with theoretical 
expectation.\,\cite{ANN} 

\begin{figure}[t]
\centering\mbox{
\epsfig{file=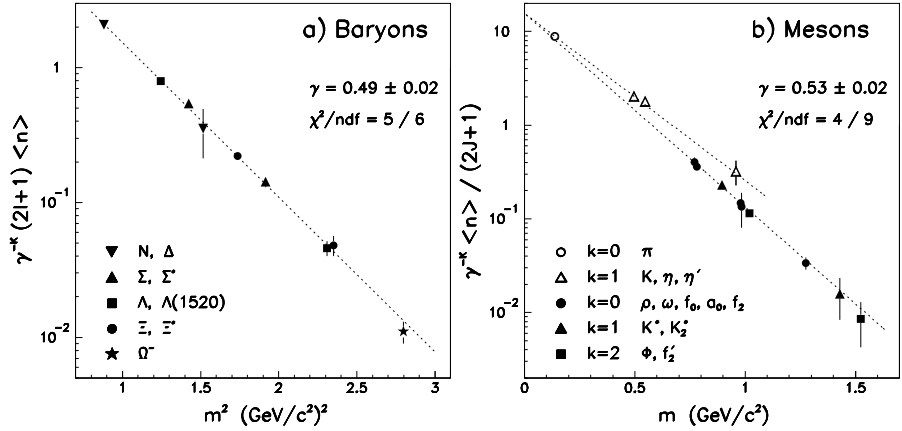,width=\textwidth}}
\caption{a) Sum of the baryon production rates of all states of an 
isomultiplet weighted by a factor $\gamma^{-k}$ as a function of the squared 
baryon mass. b) Meson production rate per spin and isospin state weighted 
by a factor $\gamma^{-k}$ as a function of the meson mass.}
\label{fig2}
\end{figure}

If the production rates are weighted by a factor $\gamma^{-k}$, a ``universal'' 
mass dependence is observed~\cite{D9} for all baryons (Fig.\,\ref{fig2}a). 
But for mesons (Fig.\,\ref{fig2}b) there are two different mass dependences; 
one for vector, tensor and scalar mesons~\cite{Panic99} and another for 
pseudoscalar mesons.\,\cite{MyPLB,Moriond} The curves in Fig.\,\ref{fig2}b show 
the result of the fit of Eq.~(\ref{mlaw}) to all meson production rates with 
the same values of $A$ and $\gamma$, but with different values of 
$b$. The observed splitting of the mass dependence of meson production rates
into two (Fig.\,\ref{fig2}b) can probably be explained by the influence 
of the spin-spin interaction between the quarks of the meson (the spins of 
quarks are parallel for vector, tensor and scalar mesons and anti-parallel for 
pseudoscalar mesons). However, there is no influence of the value and 
orientation (with respect to the net spin) of the orbital angular momentum of 
the quarks, i.e. of the spin-orbital interaction of the quarks 
(see Figs.\,\ref{fig2}a and \ref{fig2}b).

Using the values of the fitted parameter $A$ (given in Table~\ref{tab2}), the 
mass dependences of the vector-to-pseudoscalar and meson-to-baryon ratios can 
be extrapolated to the zero mass limit $m$=0, yielding:
\begin{equation}
\label{ratio1}
{{\rho^+} \over {3\,\pi^+}} \,=\, 
{{A_{\,V,T,S}} \over {A_{P}}} \,=\, 1.2\,\pm\,0.3
~~~~{\rm and}~~~~
{{\pi^+} \over {\rm p}} \,=\, 
4\cdot {{A_{M}} \over {A_{B}}} \,=\, 2.9\,\pm\,0.3,~~
\end{equation}
where $A_M = 15.3 \pm 1.2$ is the weighted average of $A_{\,V,T,S}$ and $A_P$.
The factor 4 in Eq.~(\ref{ratio1}) takes into account the fact that the baryon 
production rates fitted by Eq.~(\ref{blaw}) include charge conjugated states 
and an isospin counting factor (2$I$+1). The results (\ref{ratio1}) agree 
with the quark combinatorics model predictions (\ref{qcmratios}), although the 
latter are expected to be correct only for the directly produced hadrons.

In conclusion, using the mass dependences of the {\it total} production rates 
per hadronic Z decay of all light-flavour hadrons measured so far at LEP, 
we have shown that the $\rho$\,:\,$\pi$ and $\pi$\,:\,${\rm p}$ ratios of 
{\it ``massless''} particle production rates, obtained by extrapolating the 
empirical mass dependences (\ref{blaw}) and (\ref{mlaw}) to the zero mass 
limit $m$=0, are given by the same spin counting and quark combinatorics 
factors (\ref{qcmratios}) which are assumed in the quark combinatorics model 
for {\it direct} hadron production.
The slope splitting, observed for the dependence (\ref{mlaw}) and probably 
related to the spin-spin interaction between the quarks of the meson, and 
the difference between the $m^2$ and $m$ terms in Eqs.~(\ref{blaw}) and 
(\ref{mlaw}) lead to the violation of the vector-to-pseudoscalar and 
meson-to-baryon relations (\ref{qcmratios}) at real mass values ($m \ne 0$). 

Also the ratio of the reduced (by a spin counting factor) rates of 
``massless'' meson ($M$) and baryon ($B$) production can be written as:
\begin{equation}
\label{ratio3}
M : B ~=~ 
{{\langle n_M \rangle} \over {(2J_M+1)}} \,:\, 
{{\langle n_B \rangle} \over {(2J_B+1)}} ~=~ 
C \cdot {\lambda_{QS}} \cdot \gamma^{\,k_M-k_B}, 
\end{equation}
where $C = 2A_M/A_B$ and $\lambda_{QS} = (2J_B+1)(2I_B+1)$. 
If $C = {3 \over 2}$, Eq.~(\ref{ratio3}) coincides with the relation 
(\ref{qcmratios}) for the $\pi$\,:\,${\rm p}$\, ratio. 
The factor $\lambda_{QS}$ is an additional suppression factor.  
Eq.~(\ref{ratio3}) suggests that one ``arbitrary unit'' of the production 
rate is given to each state of each meson {\it (boson)} isomultiplet, but to 
each baryon {\it (fermion)} isomultiplet taken as a whole. Therefore, 
the $\lambda_{QS}$ can be interpreted as a fermion suppression factor 
originating from {\it quantum statistics} properties of bosons and fermions. 

The coincidence of the ratios {\it observed} for the ``massless'' particles and 
{\it expected} for the direct ones can probably be explained by the absence of 
decay processes at zero masses. However, it is not clear to us whether it is 
possible to find similar relations for the directly produced hadrons.

\bigskip
\noindent
{\it Acknowledgments.} I am grateful to W.\,Venus and F.\,Verbeure for 
the suggested improvements. I also thank V.V.\,Anisovich for useful comment.

\end{document}